\documentclass[12pt,english,preprint]{revtex4}
\usepackage[utf8]{inputenc}
\usepackage{array}
\usepackage{longtable}
\usepackage{float}
\usepackage{amsmath}
\usepackage{graphicx}
\usepackage{amssymb}
\usepackage{color}
\usepackage[FIGTOPCAP]{subfigure}

\begin{document}

\title{A Physics-Informed Deep Learning Model of the Hot Tail Runaway Electron Seed}

\author{Christopher J. McDevitt}
\affiliation{Nuclear Engineering Program, University of Florida}

\date{\today}

\begin{abstract}


A challenging aspect of the description of a tokamak disruption is evaluating the hot tail runaway electron (RE) seed that emerges during the thermal quench. This problem is made challenging due to the requirement of describing a strongly non-thermal electron distribution, together with the need to incorporate a diverse range of multiphysics processes including magnetohydrodynamic instabilities, impurity transport, and radiative losses. The present work develops a physics-informed neural network (PINN) tailored to the solution of the hot tail seed during an idealized axisymmetric thermal quench. Here, a PINN is developed to identify solutions to the adjoint relativistic Fokker-Planck equation in the presence of a rapid quench of the plasma's thermal energy. It is shown that the PINN is able to accurately predict the hot tail seed across a range of parameters including the thermal quench time scale, initial plasma temperature, and local current density, in the absence of experimental or simulation data. 
The hot tail PINN is verified by comparison with a direct Monte Carlo solution, with excellent agreement found across a broad range of thermal quench conditions.



%
%
%

\end{abstract}

\maketitle


\section{Introduction}

Runaway electrons (REs) are formed due to the force exerted by a DC electric field overcoming the Coulomb drag, thus enabling the acceleration of the electron to relativistic energies
~\cite{dwyer2014physics, breizman2019physics}.
Mechanisms through which REs may be generated can be 
separated into two broad categories. The first such category corresponds to a \emph{nonlocal} process whereby an initially low energy electron makes a discrete jump in energy due to a large-angle collision with a pre-existing high energy electron. If the energy exchanged during this collision is sufficient to scatter the initially low energy electron to high enough energy that it is accelerated by the DC electric field, this will enable the pre-existing RE population to grow. This mechanism corresponds to the avalanche mechanism of RE generation~\cite{Sokolov:1979}.



A fundamentally distinct means through which REs can be generated corresponds to the \emph{local} acceleration of electrons from the bulk electron population via a strong DC electric field. Such a local means of RE generation often provides the `seed' population of REs, which is subsequently amplified by the avalanche mechanism described above. Perhaps the most well studied example of a local process of RE generation corresponds to Dreicer acceleration~\cite{Dreicer:1959}. Here, an 
electric field 
accelerates electrons on the tail of the electron distribution with energy diffusion providing a means of filling in the regions of momentum space vacated by the accelerated electrons, thus providing a continuous source of REs.
While the Dreicer mechanism has been extensively studied~\cite{Connor:1975, hesslow2019evaluation, mcdevitt2019runaway, chacon2022asymptotic}, it is often subdominant for the high plasma densities present during a tokamak disruption
~\cite{Martin:2017, aleynikov2017generation}. A potentially larger seed mechanism active during a tokamak disruption corresponds to the hot tail mechanism of RE generation. This mechanism arises during a rapid collapse of the electron temperature, such as the thermal quench of a tokamak disruption. During such a rapid quench of the electron's thermal energy, which may be induced by the injection of impurities into the plasma for example, bulk electrons rapidly cool due to their relatively high collisionality. Energetic electrons, however, will cool on a far slower timescale thus resulting in the formation of a strongly non-thermal `hot tail' distribution of electrons~\cite{helander2004electron, smith2008hot, Stahl:2016, matsuyama2017analysis}. As the bulk plasma cools, this leads to a rise in the plasma's resistivity, resulting in the sudden growth in the strength of the electric field, and the acceleration of a fraction of the `hot tail' electron distribution to relativistic energies. 

Despite the importance of hot tail generation to fusion plasmas, an accurate treatment of this process has resisted theoretical description to date for two fundamental reasons. The first is that hot tail RE seed generation depends sensitively on the time history of the temperature profile, and thus cannot be described as a function of local parameters. While direct numerical simulations can be performed to simulate any specific temperature history, no general theory for this process is presently available. A second fundamental challenge is that part of what induces the loss of thermal energy during a disruption is the rapid transport of heat along open magnetic field lines. High energy electrons (i.e the hot tail seed) will also be susceptible to loss along such open magnetic field lines. Hence, an accurate treatment of hot tail generation requires incorporating both the specific time history of the electron temperature, along with the detailed magnetohydrodynamic (MHD) spectrum.

The present paper seeks to address the first of these challenges by providing a description of hot tail formation in the limit of an axisymmetric plasma. Since a fraction of the hot tail seed will likely be lost along open magnetic field lines, the present analysis should be viewed as an upper bound on the hot tail seed. Future work will seek to address losses via spatial transport. To carry out this analysis we will employ a deep learning framework with the aim of learning the parametric dependence of the hot tail seed. Noting that even a very simple model of a thermal quench implies a high dimensional parameter space, a purely data driven approach will require the use of a prohibitively large quantity of data to densely sample the relevant regions of parameter space. In order to overcome this limitation, we will seek to utilize physical constraints to eliminate the need to generate data to train a neural network (NN). Since such physical constraints can be sampled at an arbitrary number of points in the parameter space of the model, this will allow for the solution to be constrained even in a high dimensional parameter space. 

The present work will develop a physics-informed neural network (PINN)~\cite{raissi2019physics, karniadakis2021physics} tailored to the solution of the adjoint relativistic Fokker-Planck equation, in the presence of an evolving electric field. While PINNs allow for the incorporation of both data and physical constraints in the training of a NN, in the present analysis we will focus on the zero data limit, such that we will rely exclusively on physical constraints to identify solutions of the adjoint Fokker-Planck equation. Our aim will thus be to learn a NN based representation of the hot tail mechanism, which we will refer to as a hot tail PINN. 
The idealized model employed will involve the relativistic adjoint Fokker-Planck equation coupled to an electric field evaluated from Ohm's law during a prescribed temperature history. The hot tail seed is subsequently learned across a broad range of thermal quench times $\Delta t_{TQ}$, initial plasma temperatures $T_{init}$, and current densities $j_0$, yielding a rapid surrogate model of the hot tail seed. As a means of verifying the accuracy of the hot tail PINN's predictions, a direct Monte Carlo solution for the runaway probability function (RPF)~\cite{karney1986current} will be carried out using the RE solver RAMc~\cite{mcdevitt2019avalanche}.

The remainder of this paper is organized as follows. In Sec. \ref{sec:PCDL} the general PINN framework employed in this work is briefly reviewed. Section \ref{sec:RMTQ} describes the idealized thermal quench model employed, along with customizations of the PINN framework to the specific case of the adjoint relativistic Fokker-Planck equation. The Monte Carlo code used for verification is described in Sec. \ref{sec:VHTP}. Section \ref{sec:SMH} provides an overview of the predictions of the hot tail PINN across a broad range of thermal quench scenarios, with conclusions provided in Sec. \ref{sec:DC}.


\section{\label{sec:PCDL}Physics-Constrained Deep Learning}


A primary aim of the present paper is the development of a PINN customized to treat the adjoint relativistic Fokker-Planck equation. Before doing so, it will be useful to briefly review fundamental aspects of the PINN framework, which has emerged as a prominent example of physics-informed deep learning methods~\cite{lagaris1998artificial, karpatne2017theory, karniadakis2021physics, lusch2018deep, wang2020towards}. The present discussion will only focus on the essential concepts, where the interested reader is referred to Ref. \cite{karniadakis2021physics} for a more detailed discussion. Here, the underlying strategy is to use physical constraints (primarily PDEs) to regularize the training of a neural network. Such an approach not only provides a natural means of avoiding overfitting, thus providing a more robust interpolative tool, but also opens up the possibility of greater generalizability to unseen parameter regimes. 
A PINN in its simplest form can be expressed as~\cite{raissi2019physics, karniadakis2021physics}:
\begin{equation}
\text{Loss} = \frac{1}{N_{data}} \sum^{N_{data}}_i \left[ P_i - P \left( x_i, t_i, \lambda_i \right)\right]^2 + \frac{1}{N_{PDE}} \sum^{N_{PDE}}_i  \mathcal{R}^2 \left( x_i, t_i, \lambda_i \right)
, \label{eq:PCDL2}
\end{equation}
where $P_i$ represent data points for the quantity of interest (the RPF in the present paper), $x_i$ is a phase space coordinate (pitch or momentum, for example), $t_i$ is time, and $\lambda_i$ represent parameters of the physical system such as the time scale of the thermal quench or current density, $N_{data}$ is the number of data points used in the training of the NN, $N_{PDE}$ is the number of points where the physical constraint is sampled, and $\mathcal{R}\left( x_i \right)$ is the residual of the PDE, which penalizes deviations from the imposed physical constraint of the system.
The first term in Eq. (\ref{eq:PCDL2}) represents loss against available data, whereas the second term enforces the physical constraints.
We note that while the number of data points $N_{data}$ will be sharply limited for many challenging problems related to tokamak disruptions, the value of $N_{PDE}$ may be taken to be very large, where increasing its value will only impact the computational cost of the offline training of the model. 

For the present study we will be interested in the limit where no data is available. Our motivation will be to demonstrate the ability of PINNs to accurately learn solutions to the adjoint Fokker-Planck equation across a broad range of parameters $\lambda$ in this data free limit. As described in Sec. \ref{sec:RPFM} below, the use of a vanilla PINN such as Eq. (\ref{eq:PCDL2}) will fail to accurately evaluate the RPF in the absence of data during the highly transient conditions that are characteristic of a thermal quench. A primary motivation of the present work will thus be to develop a tailored hot tail PINN, with a loss function and neural network architecture that enables an accurate solution of the RPF across a broad range of thermal quench conditions.



\section{\label{sec:RMTQ}Reduced Model of the Thermal Quench}

\subsection{\label{sec:LDCF}Adjoint Fokker-Planck Equation}

The momentum space evolution of a relativistic population of electrons is described by the relativistic Fokker-Planck equation, i.e.
\begin{align}
\frac{\partial f_e}{\partial t} &+ \frac{1}{p^2} \frac{\partial}{\partial p} \left[ p^2 \left( -E_\Vert \xi - \frac{1+p^2}{p^2}\right) f_e \right] - \frac{\partial}{\partial \xi} \left[ \left( \frac{1-\xi^2}{p}\right) E_\Vert f_e \right] \nonumber \\
& = \frac{\nu_D}{2} \frac{\partial}{\partial \xi} \left[ \left( 1-\xi^2\right) \frac{\partial f_e}{\partial \xi}\right] + \frac{1}{p^2} \frac{\partial}{\partial p} \left( p^2 C_A \frac{\partial f_e}{\partial p}\right)
, \label{eq:TDRP1}
\end{align}
where the collisional coefficients are taken to be
\[
\nu_D = \left( 1+Z_{eff} \right)\frac{\gamma}{p^3},\quad C_A = \left( \frac{T_e}{m_e c^2}\right) \frac{\gamma^3}{p^3}
.
\]
Here, the relativistic momentum $p$ is normalized as $p\to p/ \left( m_e c\right)$, the electron's pitch is defined by $\xi \equiv p_\Vert / p$, time is normalized as $t \to t / \tau_c$, where $\tau_c \equiv 4\pi \epsilon^2_0 m^2_e c^3/ \left( e^4 n_e \ln \Lambda \right)$ is the collision time of a relativistic electron, the collisional coefficients $\nu_D$ and $C_A$ are normalized to $\tau_c$, and we have normalized the parallel electric field to the Connor-Hastie electric field $E_\Vert \to E_\Vert/E_c$, where $E_c \equiv m_e c / \left( e \tau_c \right)$~\cite{Connor:1975}. Applying a PINN directly to this equation is challenging due to the magnitude of $f_e$ varying strongly with energy, where our interest will be capturing the tail of the electron distribution which is typically orders of magnitude smaller than the bulk plasma. 

An alternate approach involves the solution of the adjoint equation~\cite{Liu:2016, zhang2017backward}
\begin{align}
\frac{\partial P \left( t \right)}{\partial t} &+ \left[ -E_\Vert \left( t \right) \xi - \frac{1+p^2}{p^2} \right]\frac{\partial P \left( t \right)}{\partial p} - \left( \frac{1-\xi^2}{p}\right) E_\Vert \left( t \right) \frac{\partial P \left( t \right)}{\partial \xi} \nonumber \\
& = -\frac{\nu_D}{2} \frac{\partial}{\partial \xi} \left[ \left( 1-\xi^2\right) \frac{\partial P \left( t \right)}{\partial \xi}\right] - \frac{1}{p^2} \frac{\partial}{\partial p} \left[ p^2 C_A \frac{\partial P \left( t\right)}{\partial p} \right]
, \label{eq:TDRP2}
\end{align}
subject to the terminal condition $P \left( t=t_{final}\right) = P_{RE} \left( p \right)$, where $P_{RE}\left( p \right) = 1$ for $p \geq p_{RE}$ and $P_{RE} \left( p \right) = 0$ otherwise. The RPF $P \left( t\right)$ represents the probability that an electron located at $\left( p,\xi \right)$ at time $t$ obtains an energy greater than $p_{RE}$ on, or before $t_{final}$~\cite{karney1986current}. Since $P$ is a probability its value will range between zero and one, and thus will be far easier to represent with a PINN compared to the electron distribution $f_e$. Equation (\ref{eq:TDRP2}) can be rewritten as an initial value problem by making the variable substitution $\tau \equiv t_{final} - t$, i.e.
\begin{align}
\frac{\partial P^\prime \left( \tau \right)}{\partial \tau} &- \left[ -E_\Vert \left( t_{final} - \tau \right) \xi - \frac{1+p^2}{p^2} \right]\frac{\partial P^\prime \left( t \right)}{\partial p} + \left( \frac{1-\xi^2}{p}\right) E_\Vert \left( t_{final} - \tau  \right) \frac{\partial P^\prime \left( \tau \right)}{\partial \xi} \nonumber \\
& = \frac{\nu_D}{2} \frac{\partial}{\partial \xi} \left[ \left( 1-\xi^2\right) \frac{\partial P^\prime \left( \tau \right)}{\partial \xi}\right] + \frac{1}{p^2} \frac{\partial}{\partial p} \left[ p^2 C_A \frac{\partial P^\prime \left( \tau\right)}{\partial p} \right]
, \label{eq:TDRP3}
\end{align}
where $P^\prime \left( \tau \right)$ is required to satisfy the initial condition $P^\prime \left( 0 \right) = P_{RE} \left( p \right)$. By making this variable change it is equivalent to beginning with 
a distribution of electrons above the RE energy $p_{RE}$, and then integrating the equations of motion backward in time to identify the 
regions of momentum space with orbits that lead to RE formation.



\subsection{\label{sec:ITQ}Axisymmetric model of the Thermal Quench}

To identify the hot tail seed, we will solve Eq. (\ref{eq:TDRP3}) in the presence of an evolving temperature and electric field. 
Specifically, the electric field will be evolved by Ohm's law, under the assumption of constant current density $j_0$, i.e.
\begin{equation}
E_\Vert = \eta j_\Vert = \eta j_0
, \label{eq:TDRP15}
\end{equation}
where we have used the normalizations $E_\Vert \to E_\Vert /E_c$, $j_0 \to a^2 j_0/I_A$ and $\eta \to I_A \eta / \left( a^2 E_c\right)$. For a Spitzer resistivity $\eta \propto 1/T^{3/2}_e$, neglecting the logarithmic dependence of the Coulomb logarithm, the electric field can then be linked to the temperature history as
\begin{equation}
E_\Vert = \left( \frac{T_{init}}{T_e}\right)^{3/2} \eta_{init} j_0 = \left( \frac{T_{init}}{T_e}\right)^{3/2} E_{init}
, \label{eq:TDRP16}
\end{equation}
where $\eta_{init}$ is the resistivity evaluated at the initial temperature $T_{init}$, and $E_{init} \equiv \eta_0 j_0$ is the initial electric field. The temperature history will be parameterized by
\begin{equation}
T_e \left( t\right) = \left( T_{init} - T_{final} \right) \exp \left( \frac{-t}{\Delta t_{TQ}}\right) + T_{final}
. \label{eq:TDRP17}
\end{equation}
Here, $T_{final}$ corresponds to the post-thermal quench temperature and $\Delta t_{TQ}$ indicates the thermal quench time scale normalized to $\tau_c$. For simplicity, we will take the plasma to be a fully ionized deuterium plasma with constant density $n_D = n_{e0}$. The Coulomb logarithm used to define the time normalization $\tau_c$ will be evaluated using the initial electron temperature $T_{init}$ and electron density $n_{e0}$.

The present model is characterized by four physics input parameters $\left( \Delta t_{TQ}, T_{init}, T_{final}, E_{init} \right)$. In addition to these four physics parameters, we will also need to specify an end time $t_{final}$, the energy range of the simulation (i.e. $p_{min}$ and $p_{max}$), and the RE energy $p_{RE}$. The solution can be shown to be insensitive to these parameters so long as they are chosen to be sufficiently large (or small for the case of $p_{min}$). 


\subsection{\label{sec:RPFM}Hot Tail PINN}

A property of NNs that has led to their widespread use across a range of applications is the impressive expressivity of NNs~\cite{hornik1989multilayer}. When applying NNs to scientific problems, however, it is often preferable to restrict the range of functions that can represented by the NN to those that are consistent with the physical problem of interest. In so doing, the space of available solutions is drastically reduced leading to a far more robust training process~\cite{lu2021physics}. Our aim in this subsection will be to impose as hard constraints in the NN architecture several properties of the RPF. Specifically, we will restrict the range of values of the RPF to be between zero and one, enforce the initial condition to be $P^\prime \left( t=0 \right) = P_{RE} \left( p \right)$, and enforce the boundary condition $P_{RE} \left( p=p_{min}\right) = 0$ as hard constraints. In addition, we will enforce the condition $P^\prime \left( p=p_{max}, \xi=\xi_{min}\right) = 1$, where $\xi_{min} = -1$. While this condition is not strictly true for all possible thermal quench scenarios, will choose $p_{max}$ to be sufficiently large such that it will be satisfied for all cases considered in this paper. In practice, forcing $P^\prime$ to be one at this location helps ensure robust convergence of the PINN. 


The means through which we will enforce these conditions will be by introducing additional layers to the NN.
Specifically, defining the output of the NN by $P_{NN}$, we will introduce a layer of the form
\begin{equation}
\bar{P}^\prime \left( p,\xi,t ;\mathbf{\lambda}\right) = P_{RE} \left( p\right) + \left( \frac{p-p_{min}}{p_{max}-p_{min}} \right) \left( \frac{t-t_{min}}{t_{max}-t_{min}} \right) \left( \frac{p_{max}-p \Xi \left( \xi\right)}{p_{max}-p_{min}} \right) P_{NN} \left( p,\xi,t ;\mathbf{\lambda} \right)
. \label{eq:RPFM1}
\end{equation}
To ensure the RPF has a range between zero and one we will pass this the final layer defined by
\begin{equation}
P^\prime \left( p,\xi,t;\mathbf{\lambda} \right) = \tanh \left( \frac{{\bar{P}^\prime \left( p,\xi,t,\mathbf{\lambda}\right) }^2}{\Delta_{prob}^2} \right)
, \label{eq:RPFM2}
\end{equation}
where $\mathbf{\lambda}$ indicates the parameters of the hot tail model $\left( \Delta t_{TQ}, T_{init}, T_{final}, j_0 \right)$, we will choose $\Delta^2_{prob} \ll 1$, and we have defined the function
\begin{equation}
\Xi \left( \xi \right) \equiv \exp \left( -\frac{ \left( \xi - \xi_{min}\right)^2 }{\Delta \xi^2} \right)
. \label{eq:LPD2}
\end{equation}
These output layers can be verified to constrain $P^\prime$ to have a range between zero and one, forces the solution to vanish at $p=p_{min}$, exactly enforces the initial condition $P^\prime = P_{RE}$, and satisfies the property $P^\prime \left( p=p_{max}, \xi=\xi_{min}\right) = 1$ which is true for sufficiently large $p_{max}$.



Rather than employing a step function for the initial condition $P_{RE} \left( p \right)$, we will introduce a modest smoothing of this initial condition as follows
\begin{equation}
P_{RE} \left( p \right) = \frac{1}{2} \left[ 1 - \tanh \left( \frac{p_{RE}-p}{d p}\right)\right]
. \label{eq:LPD3}
\end{equation}
Noting that 
the PINN will be learning deviations from $P_{RE}$ for $t>0$, by using a smooth form of $P_{RE}$ the PINN does not need to learn a sharp discontinuity.

For all of the cases considered below we will take the various parameters characterizing these additional output layers to be fixed. Specifically, we will take $\Delta \xi = 0.5$, $dp = 0.1 p_{max}$, $\Delta_{prob} = 0.25$ and $p_{RE}= 0.8 p_{max}$. While the value for $dp$ may seem somewhat large, since $P_{RE}$ will be squared in Eq. (\ref{eq:LPD3}), and then evaluated inside a $\tanh$, this choice of $dp$ can be verified to lead to a sharp transition between zero and one. For all cases considered in this work we will take $p_{max}\approx 1.7$, or a maximum energy of $500\;\text{keV}$. With regard to the NN, we will employ a simple feedforward fully connected NN, with six hidden layers and 64 neurons per hidden layer. Finally we note that we have made use of the DeepXDE library~\cite{lu2021deepxde} for the studies carried out below, where all source code will be made available through GitHub after acceptance of this article.

\section{\label{sec:VHTP}Verification of Hot Tail PINN: Monte Carlo Runaway Electron Simulations}

\begin{figure}
\begin{centering}
\subfigure[]{\includegraphics[scale=0.5]{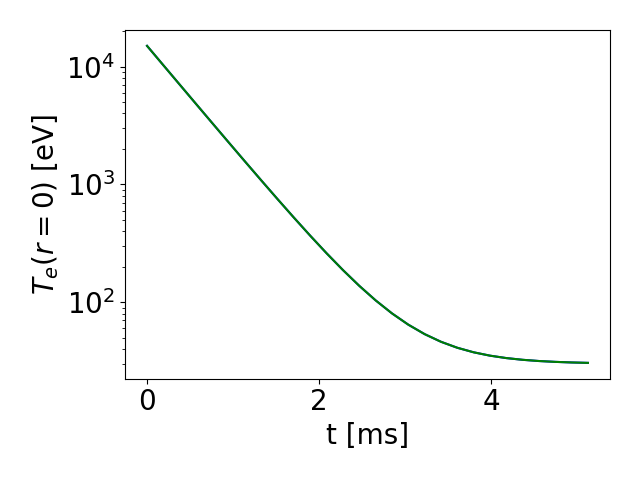}}
\subfigure[]{\includegraphics[scale=0.5]{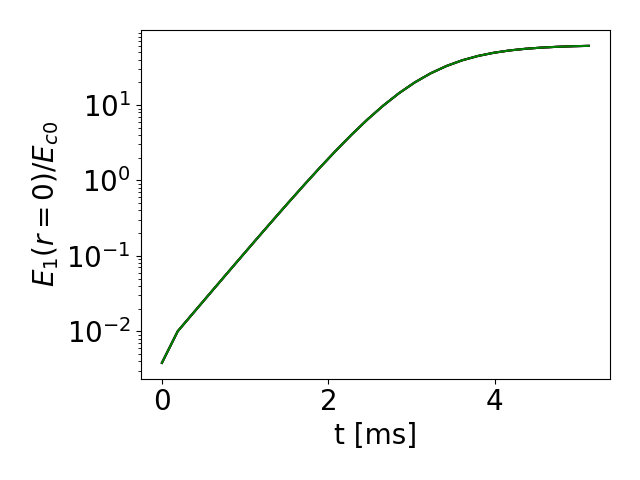}}
\par\end{centering}
\caption{Example time history of the electron temperature [panel (a)] and electric field [panel (b)] used in the RAMc simulations. The thermal quench time was taken to be $\Delta t_{TQ}=0.5\;\text{ms}$, $j_0 = 1\;\text{MA}/\text{m}^2$, and the initial and final temperatures are given by $T_{init} = 15\;\text{keV}$ and $T_{final} = 30\;\text{eV}$, respectively.}
\label{fig:VHTP1}
\end{figure}

The accuracy of the hot tail PINN will be verified by performing direct simulations of hot tail formation using the RunAway Monte carlo (RAMc) code. This particle based RE solver evaluates the guiding center motion of relativistic electrons including both small and large-angle collisions~\cite{mcdevitt2019avalanche}. The inductive electric field within RAMc is updated self-consistently by a flux diffusion equation accounting for the current carried by REs
~\cite{mcdevitt2022runaway}. In order to provide direct comparisons with predictions of the hot tail PINN and RAMc, several simplifications to the RAMc solver will be introduced. In particular, the collisional coefficients and the flux diffusion equation used in RAMc have been modified to be consistent with Secs. \ref{sec:LDCF} and \ref{sec:ITQ} above. Specifically, the large-angle collision operator has been disabled such that we will only describe seed RE formation. Also, the coupling of the RE current back to the flux diffusion equation has also been disabled to better approximate the simplified Ohm's law used in the hot tail PINN model [i.e. Eq. (\ref{eq:TDRP16})]. While the plasma current will evolve slightly during the RAMc simulations, we will use a relatively high post thermal quench plasma temperature ($T_e = 30\;\text{eV}$), which implies a modest resistivity, to ensure the plasma current density undergoes little change during the short timescale of the thermal quench. Furthermore, the marker particles are all initialized at $r\approx 0$ such that the impact of tokamak geometry on RE formation will be negligible. This enables comparisons between this toroidal solver and the slab RE model described by Eq. (\ref{eq:TDRP3}). A large ITER-like device with a minor radius of $a=200\;\text{cm}$, inverse aspect ratio of $\epsilon = 1/3$, and an on-axis magnetic field of $B_0 = 5.3\;\text{T}$ is selected when performing the RAMc simulations to ensure electrons remain well confined to their initial flux surfaces. In addition, since low energy particles have a vanishingly small probability of running away, an absorbing boundary will be implemented at an energy of $100\;\text{eV}$.

\begin{figure}
\begin{centering}
\subfigure[]{\includegraphics[scale=0.5]{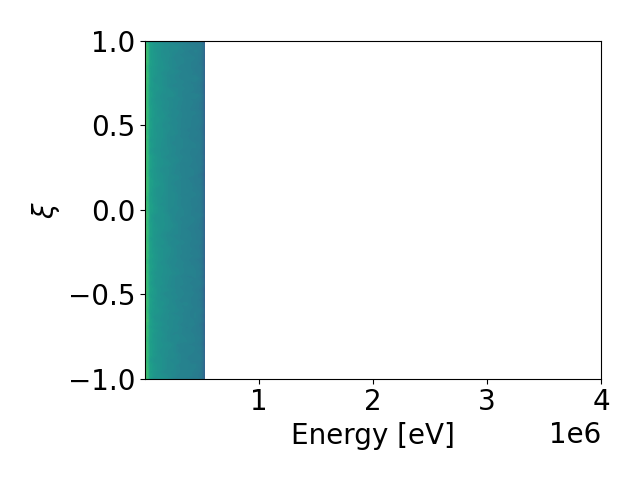}}
\subfigure[]{\includegraphics[scale=0.5]{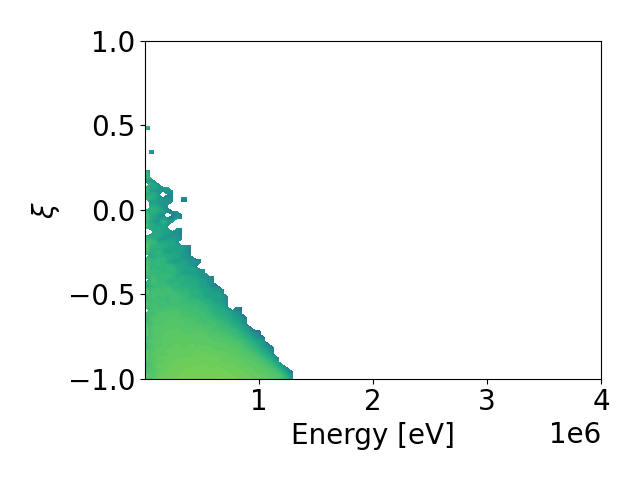}}
\subfigure[]{\includegraphics[scale=0.5]{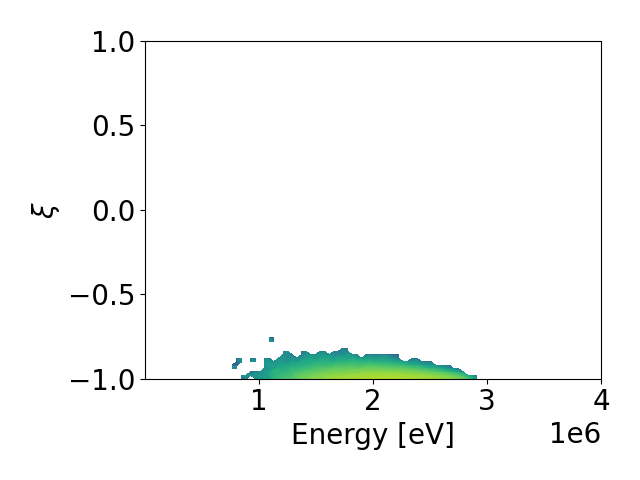}}
\subfigure[]{\includegraphics[scale=0.5]{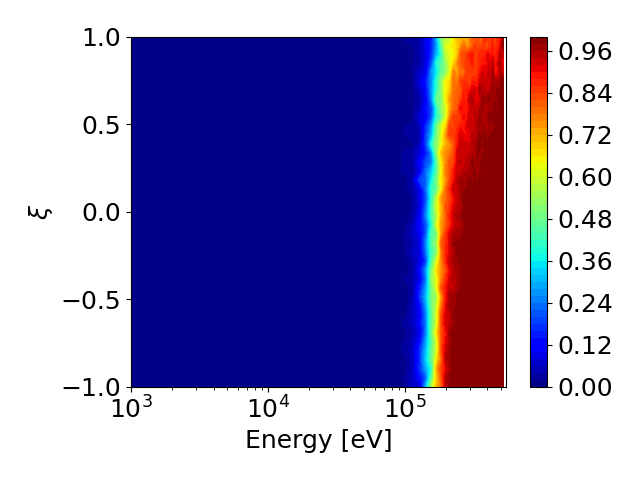}}
\par\end{centering}
\caption{Particle distributions in pitch and energy at $t=0$ [panel (a)], $t\approx3.80\;\text{ms}$ [panel (b)], and $t\approx4.94\;\text{ms}$ [panel (c)]. The RPF is shown in panel (d). The thermal quench time was taken to be $\Delta t_{TQ}=0.5\;\text{ms}$, $j_0 = 1\;\text{MA}/\text{m}^2$, and the initial and final temperatures are given by $T_{init} = 15\;\text{keV}$ and $T_{final} = 30\;\text{eV}$, respectively.}
\label{fig:VHTP2}
\end{figure}

With the above simplifications to the RAMc code, 
direct Monte Carlo calculations of the RPF will be employed to verify the predictions of the hot tail PINN.
This is done by initializing marker particles according to a uniform random distribution in momentum space at $t=0$. All electrons will have an initial spatial location taken to be $r\approx 0$. By then evaluating which electrons are able to achieve an energy of $p_{RE}$ or greater after $t=t_{final}$, this allows $P^\prime \left( \tau = t_{final}\right)$ to be directly evaluated from the particle data. An example solution is shown in Figs. \ref{fig:VHTP1} and \ref{fig:VHTP2}. Here, 100,000 marker particles are used to evaluate the RPF, where for this example the parameters were assumed to be $\Delta t_{TQ}=0.5\;\text{ms}$, $T_{init} = 15\;\text{keV}$, $T_{final} = 30\;\text{eV}$, and $E_0$ is evaluated for a fully ionized deuterium plasma with $n_D=10^{14}\;\text{cm}^{-3}$ and $j_0 = 1\;\text{MA/m}^2$. The time history of the on-axis electric field and temperature are shown in Fig. \ref{fig:VHTP1}, with the resulting RPF and particle distribution shown in Fig. \ref{fig:VHTP2}. Here, the rapid drop in the temperature leads to the formation of a large electric field with a maximum value of roughly $60$ times the Connor-Hastie threshold. As evident from the particle distributions, after roughly ten thermal quench times ($\approx 5\;\text{ms}$), all of the particles have either been absorbed by the low energy boundary, or been accelerated past the RE energy $p_{RE}$ of $\approx 300\;\text{keV}$. Noting that the particles that run away have an energy well in excess of $p_{RE}$, implies that there is little sensitivity to the specific choice of $p_{RE}$, as long as the RE energy is below an MeV. From the form of the RPF [Fig. \ref{fig:VHTP2}(d)], it is apparent that electrons whose initial energy is less than $\sim100\;\text{keV}$ have a negligible chance of running away, implying that the dominant contribution to the RE seed will be from electrons with energies far in excess of the thermal energy. The magnitude of the hot tail seed can be evaluated from this RPF by evaluating the integral
\begin{equation}
n_{RE} = \int d^3 p f^{MJ}_e P \left( p,\xi \right)
, \label{eq:VHTP1}
\end{equation}
where $P \left( p,\xi \right)$ is the RPF indicated in Fig. \ref{fig:VHTP2}(d), and $f^{MJ}_e$ is a Maxwell-J\"{u}ttner distribution evaluated using the initial electron temperature $T_{init}$ and electron density $n_{e0}$. Physically, $P \left( p,\xi \right)$ indicates the probability that an electron initially at a given momentum $\left( p, \xi\right)$ is able to run away, thus by integrating with the initial electron distribution we will be able to compute the number of REs. While Monte Carlo solutions are computationally intensive, 
a small number of RPFs of the form shown in Fig. \ref{fig:VHTP2}(d) will be used to verify the hot tail PINN solution. We emphasize that this data set will be used for verification, and not included in the training of the hot tail PINN.

\section{\label{sec:SMH}Predictions of the Hot Tail PINN}

\subsection{\label{sec:HTS}Runaway Probability Function Evolution During the Thermal Quench}

\begin{figure}
\begin{centering}
\subfigure[]{\includegraphics[scale=0.5]{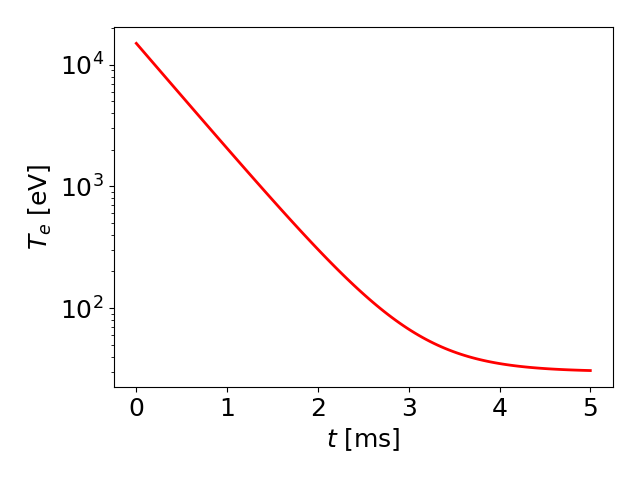}}
\subfigure[]{\includegraphics[scale=0.5]{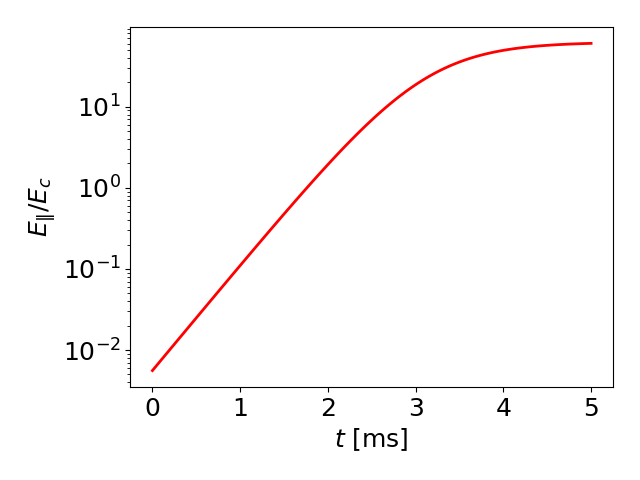}}
\par\end{centering}
\caption{Example time history of the electron temperature [panel (a)] and electric field [panel (b)]. The thermal quench time was taken to be $\Delta t_{TQ}=0.5\;\text{ms}$, $j_0 = 1\;\text{MA}/\text{m}^2$, and the initial and final temperatures are given by $T_{init} = 15\;\text{keV}$ and $T_{final} = 30\;\text{eV}$, respectively.}
\label{fig:HTS1}
\end{figure}

We will be interested in evaluating the RPF from Eq. (\ref{eq:TDRP3}) for the time dependent temperature and electric fields described in Sec. \ref{sec:ITQ}. 
For this first case, we will take $\left( T_{init}, T_{final}, E_0 \right)$ to be fixed, and seek to learn the solution as a function of $\Delta t_{TQ}$. Specifically, we will set $T_{init} = 15\;\text{keV}$ and $T_{final} = 30\;\text{eV}$, with $E_0$ evaluated for a quasi-neutral deuterium plasma with $n_{D}=10^{14}\;\text{cm}^{-3}$ and $j_0=1\;\text{MA}/\text{m}^2$. The hot tail PINN will be trained for thermal quench times ranging from $0.25\;\text{ms}$ to $2\;\text{ms}$. The total time of each simulation is taken to be ten times the thermal quench time. The time evolution of both the temperature and electric field are shown in Fig. \ref{fig:HTS1} for $\Delta t_{TQ}= 0.5\;\text{ms}$. Comparing with the time history used in the RAMc simulation for identical parameters (Fig. \ref{fig:VHTP1}), the two sets of fields are in good agreement, indicating that we will be able to perform a direct comparison between predictions of the hot tail PINN and the Monte Carlo simulations.



\begin{figure}
\begin{centering}
\subfigure[]{\includegraphics[scale=0.5]{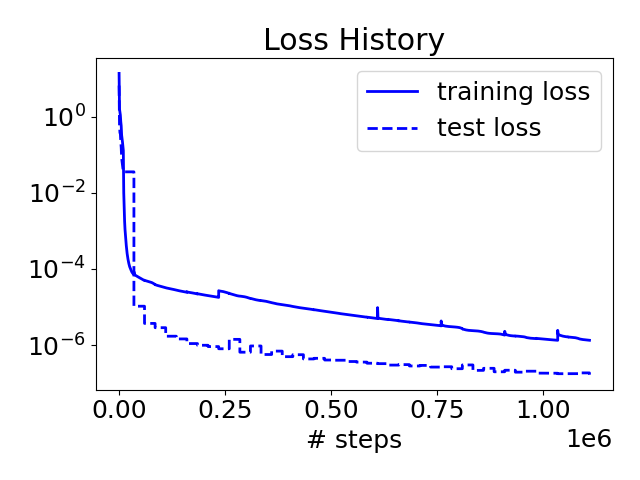}}
\subfigure[]{\includegraphics[scale=0.5]{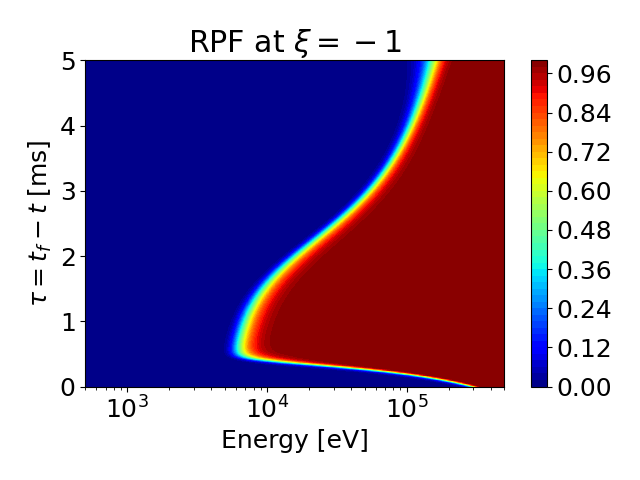}}
\par\end{centering}
\caption{Loss history [panel (a)] and RPF at $\xi=-1$ [panel (b)]. 10,000 iterations were performed with the ADAM optimizer, followed by over a million iterations performed with the L-BFGS optimizer. The thermal quench time was taken to be $\Delta t_{TQ}=1\;\text{ms}$, $j_0 = 1\;\text{MA}/\text{m}^2$, and the initial and final temperatures are given by $T_{init} = 15\;\text{keV}$ and $T_{final} = 30\;\text{eV}$, respectively.}
\label{fig:HTS1sub1}
\end{figure}

The loss history of the RPF trained with these evolving temperature and electric fields is shown in Fig. \ref{fig:HTS1sub1}(a).  At increments of twenty-five thousand steps additional training points are added at locations where the residual of the PDE is maximal, causing the periodic spikes in the training loss evident in Fig. \ref{fig:HTS1sub1}(a). Due to the sharp variation of the RPF when the electric field is maximal [i.e. for $\tau\equiv t_{final}-t \approx 0$, see Fig. \ref{fig:HTS1sub1}(b)], additional training points have also been added to the initial $10$\% of the simulation, i.e. from $\tau =0$ to $\tau = 0.5\;\text{ms}$ for the case shown in Fig. \ref{fig:HTS1sub1}(b). These two differences in the training versus test point distribution lead to a significant deviation in the training versus test loss, with the 
training loss dropping to a value of approximately $1.3\times 10^{-6}$ after $\sim 1,000,000$ iterations, with a test loss of roughly $1.7\times 10^{-7}$.
The approximate seven orders of magnitude drop of the test loss indicates that the hot tail PINN was able to identify a well converged solution.

\begin{figure}
\begin{centering}
\subfigure[]{\includegraphics[scale=0.5]{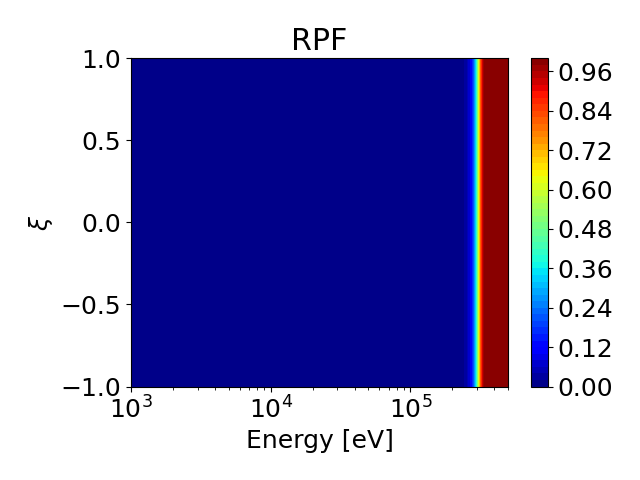}}
\subfigure[]{\includegraphics[scale=0.5]{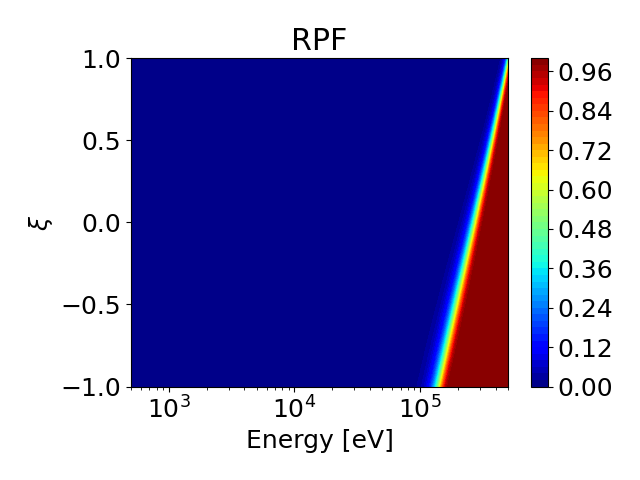}}
\subfigure[]{\includegraphics[scale=0.5]{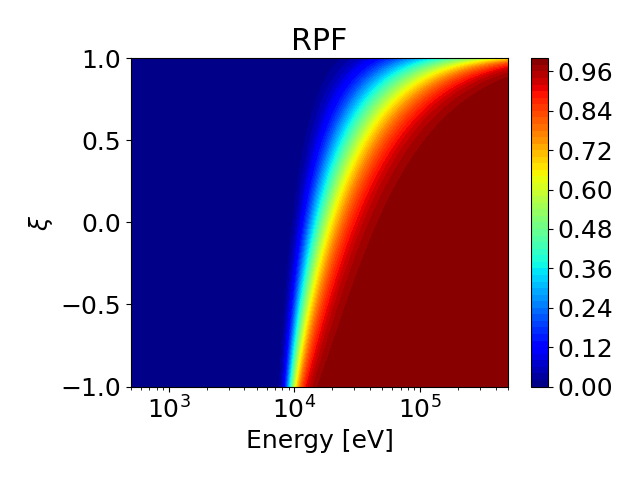}}
\subfigure[]{\includegraphics[scale=0.5]{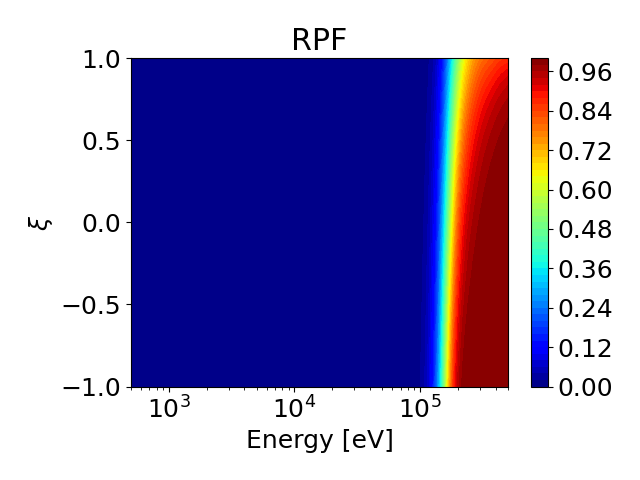}}
\par\end{centering}
\caption{RPF at $\tau=0$ [panel (a)], $\tau=0.15\;\text{ms}$ [panel (b)], $\tau=1.5\;\text{ms}$ [panel (c)] and $\tau=5\;\text{ms}$ [panel (d)]. The thermal quench time was taken to be $\Delta t_{TQ}=0.5\;\text{ms}$, with the initial and final temperatures given by $T_{init} = 15\;\text{keV}$ and $T_{final} = 30\;\text{eV}$, respectively.}
\label{fig:HTS2}
\end{figure}

Time slices of the RPF computed from the hot tail PINN are indicated in Fig. \ref{fig:HTS2}. The first time slice [Fig. \ref{fig:HTS2}(a)] indicates the terminal RPF, where $P^\prime \left( \tau = 0\right) = P_{RE}$. As indicated in Sec. \ref{sec:RPFM} the terminal RPF is taken to be a modestly smoothed step function.
At $\tau=0$ (i.e. $t=t_{final}$), the electric field is at its maximum value, which noting the backward nature of Eq. (\ref{eq:TDRP3}), implies electrons with $\xi=-1$ will move to lower energy, whereas those with $\xi=1$ will increase their energy. This results in the RPF front for electrons shifting downward in energy for $\xi=-1$ and moving to higher energy for $\xi=1$ [compare Fig. \ref{fig:HTS2}(a) with \ref{fig:HTS2}(b)]. By $\tau = 1.5\;\text{ms}$, the RPF takes on a form similar to the steady state solution~\cite{Liu:2017}. As the system evolves further the electric field decreases in magnitude, resulting in the RPF front shifting to higher energy. At $\tau = t_{final}$ the RPF front has shifted to approximately $E \approx 150\;\text{keV}$. This last RPF distribution indicates the probability of an electron running away at or before $t_{final}$, and is thus the appropriate value to use when evaluating the hot tail seed. Comparing this RPF with the RPF evaluated by RAMc shown in Fig. \ref{fig:VHTP2}(d) above, good agreement is evident between these two results indicating that the hot tail PINN was able to accurately learn the RPF.


\subsection{\label{sec:nRE}Hot Tail Seed}

\subsubsection{\label{sec:DTQ}Dependence on the Timescale of the Thermal Quench}

\begin{figure}
\begin{centering}
\subfigure[]{\includegraphics[scale=0.5]{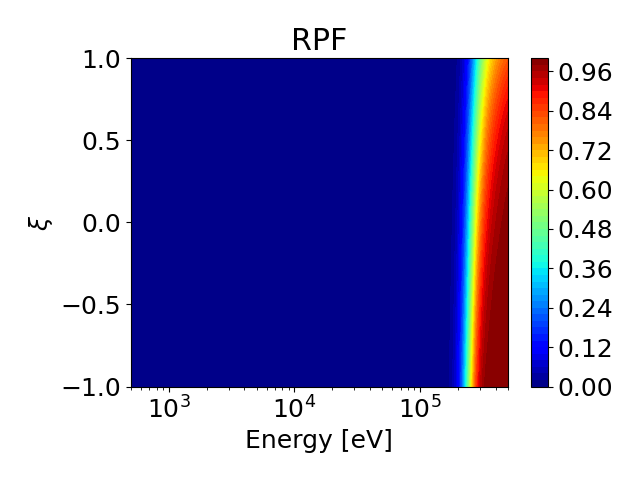}}
\subfigure[]{\includegraphics[scale=0.5]{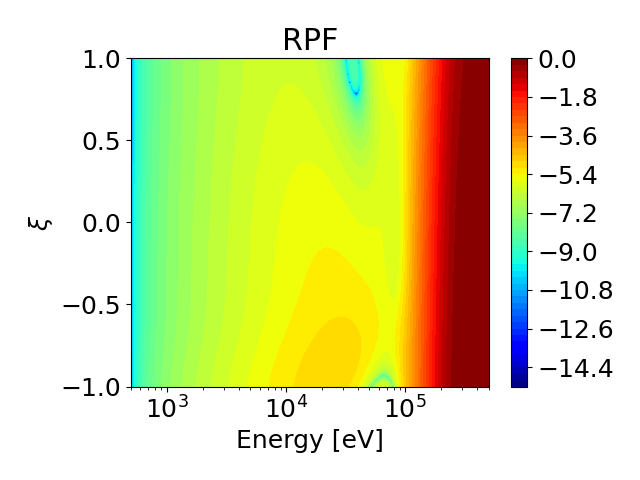}}
\subfigure[]{\includegraphics[scale=0.5]{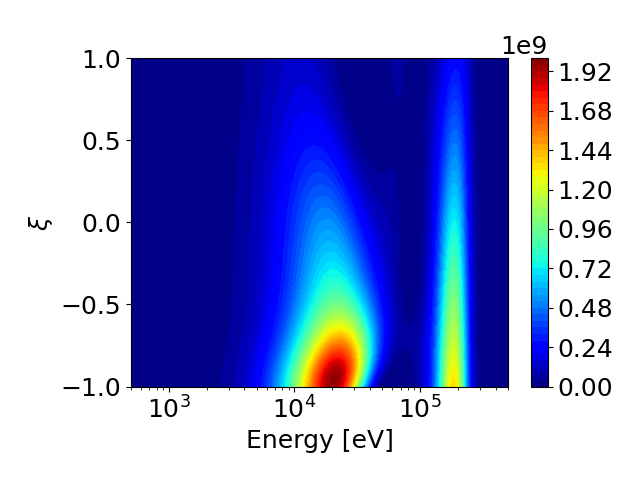}}
\subfigure[]{\includegraphics[scale=0.5]{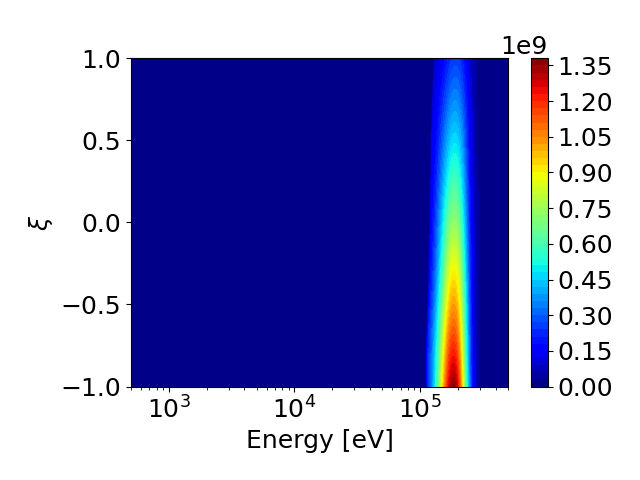}}
\par\end{centering}
\caption{RPF at $\tau=t_{final}$ [panel (a)], log scale plot of the RPF at $\tau=t_{final}$ [panel (b)], integrand of Eq. (\ref{eq:nRE1}) [panel (c)], filtered integrand of Eq. (\ref{eq:nRE2}) [panel (d)]. The thermal quench time was taken to be $\Delta t_{TQ}=0.5\;\text{ms}$, with the initial and final temperatures given by $T_{init} = 15\;\text{keV}$ and $T_{final} = 30\;\text{eV}$, respectively.}
\label{fig:nRE1}
\end{figure}

Our first task will be to use the RPF evaluated in Sec. \ref{sec:HTS} to evaluate the hot tail seed.
Taking the electron distribution before the thermal quench to be a Maxwell-J\"{u}ttner distribution $f^{MJ}_e$ with a temperature of $T_{init}$ and density $n_{e0}$, the number of hot tail electrons will then be given by:
\begin{equation}
n_{RE} = \int d^3 p f^{MJ}_e P^\prime \left( \tau=t_{final} \right)
, \label{eq:nRE1}
\end{equation}
where $P^\prime \left( \tau=t_{final} \right)$ will be evaluated using the hot tail PINN.
A complication in performing this analysis arises due to the Maxwell-J\"{u}ttner distribution being peaked at low energies, and then decaying exponentially with energy, 
such that a small error in the RPF at low energies could lead to a spurious contribution to $n_{RE}$. This complication is illustrated in Fig. \ref{fig:nRE1} for $\Delta t_{TQ} = 1\;\text{ms}$. The RPF at $\tau=t_{final}$ is shown in Fig. \ref{fig:nRE1}(a), where the RE transition region is centered around $E \approx 260\;\text{keV}$, with the RPF having a small value for energies less than $\approx 200\;\text{keV}$. However, considering the integrand that enters Eq. (\ref{fig:nRE1}) [Fig. \ref{fig:nRE1}(c)] two maximums are evident. The first is centered at an energy of approximately $190\;\text{keV}$, and is peaked near the $\xi=-1$ axis. This corresponds to the peak associated with hot tail generation. In particular, while the RPF has a modest value at this energy, noting that a Maxwell-J\"{u}ttner distribution decays exponentially with energy, the dominant contribution to the hot tail seed is expected to occur at the transition front of the RPF. We will need to capture this maximum of the integrand to evaluate the hot tail seed.

A second maximum of the hot tail integrand is also evident from Fig. \ref{fig:nRE1}(c). This maximum spans energies between $\sim8\;\text{keV}$ up to values modestly under $\sim40\;\text{keV}$. There is no physical mechanism that would generate REs from this region of momentum space. The origin of this peak can instead be tracked to a small inaccuracy of the RPF at low energy. Considering a log scale plot of the RPF [Fig. \ref{fig:nRE1}(b)], the shape of the RPF for energies less than $100\;\text{keV}$ is slightly distorted, with a maximum centered at $\sim 20\;\text{keV}$ evident. This unphysical feature is due to the RPF being exceptionally small in this region, with the artificial maximum having a magnitude of $\sim10^{-4}$, such that the residual of Eq. (\ref{eq:TDRP3}) does not heavily weigh this region. 

In order to remove this low energy unphysical contribution to the hot tail integrand, we will choose a weighting scheme for the loss function that enables regions with $P^\prime \ll 1$ to be strongly weighted, along with  post-process
the predictions of the hot tail PINN. With regard to the former approach, to increase the weight placed on regions where the RPF is small, we have defined the loss function to be of the form:
\begin{equation}
\text{Loss} = \frac{1}{N_{PDE}} \sum^{N_{PDE}}_i \left[ \left( \frac{1}{\delta+P^\prime \left( p_i, \xi_i, \tau_i, \mathbf{\lambda}_i \right)} \right) \left( \frac{p^2_i}{1+p^2_i} \right) \mathcal{R}_i \right]^2
. \label{eq:DTQ1}
\end{equation}
Here, $\mathcal{R}_i$ corresponds to the residual of Eq. (\ref{eq:TDRP3}), the factor $p^2_i / \left( 1+p^2_i\right)$ removes the divergence of the pitch-angle scattering operator at low energy, and the factor $1/\left( \delta + P^\prime \right)$, increases the weight of regions with small values of $P^\prime$. A loss function of this form was used when training the RPF shown in Fig. \ref{fig:nRE1}, with $\delta$ chosen to be $\delta = 0.05$, which enabled the 
hot tail PINN to learn the RPF with sufficient accuracy such that a clear separation between the physical and unphysical contributions to the hot tail integrand is evident. Smaller values of $\delta$ could be used to further separate the physical and unphysical contributions to the hot tail integrand, however, the $\delta$ used in the present example was sufficient to enable a simple post processing algorithm to remove the spurious contribution from the integrand.

The first step of the post processing algorithm will be to limit the integration region used in Eq. (\ref{eq:nRE1}) 
to energies three times greater than the thermal energy. In so doing the very low energy region, where spurious features are most likely, will be removed. For the example indicated in Fig. \ref{fig:nRE1} this filter will remove the contribution to the integrand for electrons with energies less than $45\;\text{keV}$ and is thus sufficient to nearly completely remove the unphysical contribution. By limiting ourselves to electrons whose initial energy is greater than three times the thermal energy the maximum hot tail seed that the model will be able to predict is given by $n_{RE}/n_{e0} \approx 0.12$.
However, noting that a relativistic electron population with $\xi=-1$ and a magnitude of $n_{RE}/n_{e0} \sim 10^{-4}$ is sufficient to carry most, if not all, of the plasma current, the prediction of a RE seed comparable to $n_{RE}/n_{e0} \sim 10^{-1}$ would be unphysical. In particular, for cases where particularly large RE seeds are present, this would lead to a modification of the Ohm's law defined by Eq. (\ref{eq:TDRP16}).

Noting that the maximum contribution to the hot tail integrand increases as $\Delta t_{TQ}$ is increased, adjusting the lower bound on the integration will not be sufficient for very slow thermal quenches. 
An additional filter will be introduced by removing regions from the integrand where the RPF is smaller than a critical value, indicated by $\Delta P \ll 1$. In so doing, the predicted hot tail seed will be computed only using regions where the RPF has a value greater than $\Delta P$, which can be chosen to exclude unphysical contributions such as those indicated in Fig. \ref{fig:nRE1}(b). As a final filter we will use the residual of the PDE to remove regions where the relative accuracy of the solution is low. This is accomplished by only including contributions where the residual of the PDE $R_{PDE}$ normalized to $P^\prime$ is greater than a threshold value $\Delta R_{thres}$. Specifically, defining the normalized residual $R_{norm}\equiv R_{PDE} / P^\prime$, we will require the integrand of Eq. (\ref{eq:nRE1}) to vanish when $\left| R_{norm} \right| > \Delta R_{thres}$. 
These filters will be implemented by defining the function
\[
\chi \left( P^\prime \right) = \left[ 1 - \exp \left( -\frac{{P^\prime}^{10}}{\Delta P^{10}} \right) \right] \exp \left( -\frac{{R}^{10}_{norm}}{\Delta R^{10}_{thres}} \right)
,
\]
such that the filtered integrand for Eq. (\ref{eq:nRE1}) becomes
\begin{equation}
n^{filter}_{RE} = \int d^3 p f^{MJ}_e P^\prime \left( \tau=t_{final} \right) \chi \left( P^\prime \right) 
. \label{eq:nRE2}
\end{equation}
The integrand of Eq. (\ref{eq:nRE2}) is shown in Fig. \ref{fig:nRE1}(d) for the parameters $\Delta P = 10^{-4}$ and $\Delta R_{thres}=1$. It is evident that the spurious maximum in the integrand has been removed, whereas the maximum associated with hot tail generation is unaffected.

\begin{figure}
\begin{centering}
\subfigure[]{\includegraphics[scale=0.5]{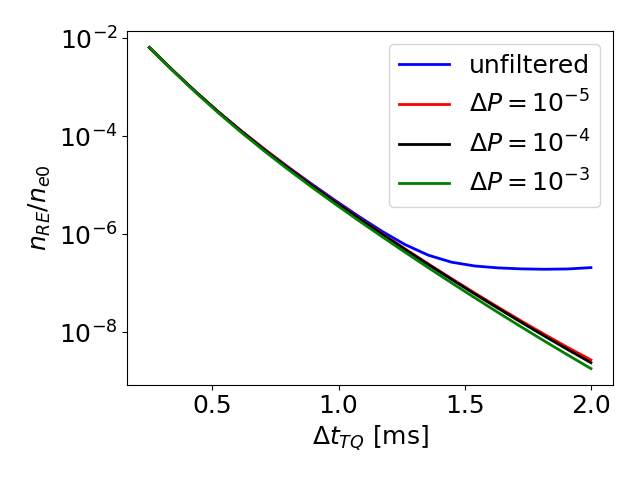}}
\subfigure[]{\includegraphics[scale=0.5]{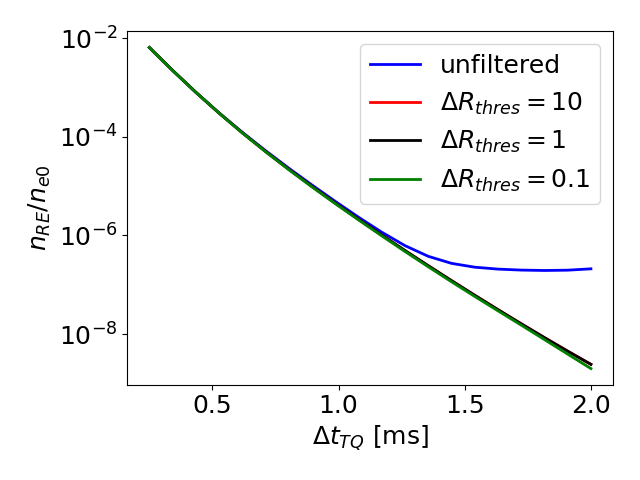}}
\par\end{centering}
\caption{Predicted hot tail seed for different values of $\Delta P$ [panel (a)] and different values of $R_{thres}$ [panel (b)]. For panel (a), $R_{thres}$ was set to $1$ whereas for panel (b) $\Delta P$ was set to $10^{-4}$. The integration region was taken to be between $E_{min}=3T_{init} = 45\;\text{keV}$ and $E_{max} = 500\;\text{keV}$.}
\label{fig:nRE3}
\end{figure}

To determine the sensitivity of the predicted hot tail seed to the choice of $\Delta P$ and $\Delta R_{thres}$, the predicted hot tail seed versus the thermal quench time is shown in Fig. \ref{fig:nRE3} for several values of $\Delta P$ and $\Delta R_{thres}$.
It is evident that for rapid thermal quenches ($\Delta t_{TQ} \lesssim 1\;\text{ms}$) all cases are in approximate agreement with each other, implying that the only filtering required is to set the low energy limit of the hot tail integrand to $3T_{init}$. However, for slower thermal quenches, the unfiltered result deviates from the filtered cases.
This is due to the transition region of the RPF shifting to higher energies for larger $\Delta t_{TQ}$, allowing the integrand to be more easily polluted by small inaccuracies in the RPF. The number of predicted hot tail REs is weakly sensitive to the values of $\Delta P$ and $\Delta R_{thres}$ used, where the percent difference is largest for slow thermal quenches. For the remainder of this paper we will use $\Delta P = 10^{-4}$ and $\Delta R_{thres}=1$. We note that for all the cases treated in this paper, a filter based on excluding contributions to the integrand with a value of the RPF below $10^{-4}$ is sufficient to remove spurious contributions to the integrand. We have introduced the two other approaches, limiting the range of integrand and residual based filtering, with an eye toward future studies where higher dimensional hot tail PINNs will be employed, and a more aggressive post-processing procedure may be needed.

\subsubsection{\label{sec:GS}General Solution}

\begin{figure}
\begin{centering}
\includegraphics[scale=0.45]{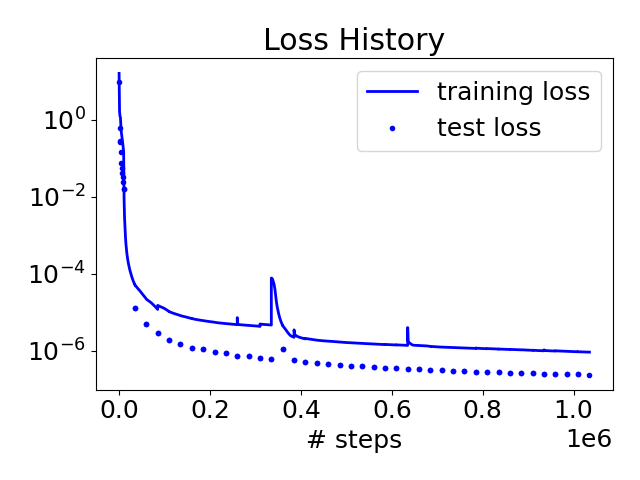}
\par\end{centering}
\caption{Loss history for a 6-D model. The model was trained for $\Delta t_{TQ}$ between $0.25\;\text{ms}$ and $2 \;\text{ms}$, $T_{init}$ between $10\;\text{keV}$ and $20\;\text{keV}$, and current densities between $j_0=0.5\;\text{MA}$ and $j_0=1.5\;\text{MA}$. The other parameters of the model were taken to be $T_{final} = 30\;\text{eV}$ and $Z_{eff}=1$.}
\label{fig:GS1}
\end{figure}

\begin{figure}
\begin{centering}
\subfigure[]{\includegraphics[scale=0.5]{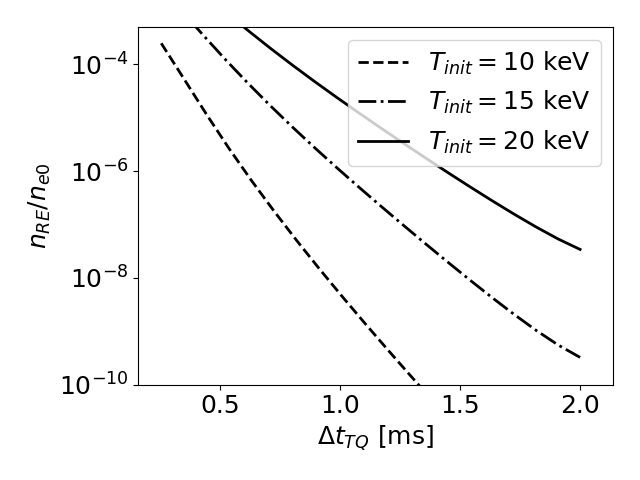}}
\subfigure[]{\includegraphics[scale=0.5]{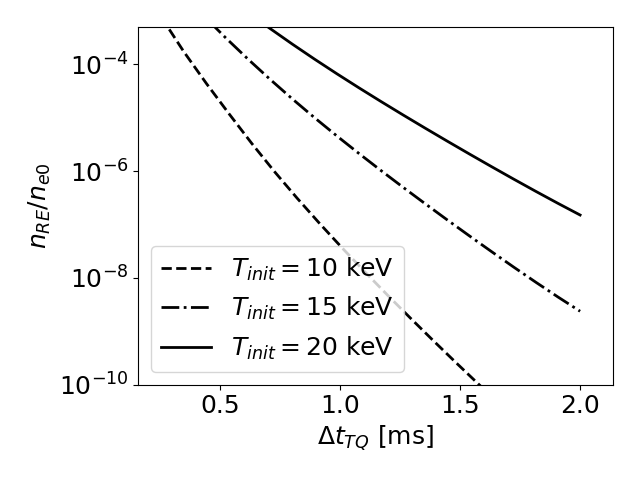}}
\subfigure[]{\includegraphics[scale=0.5]{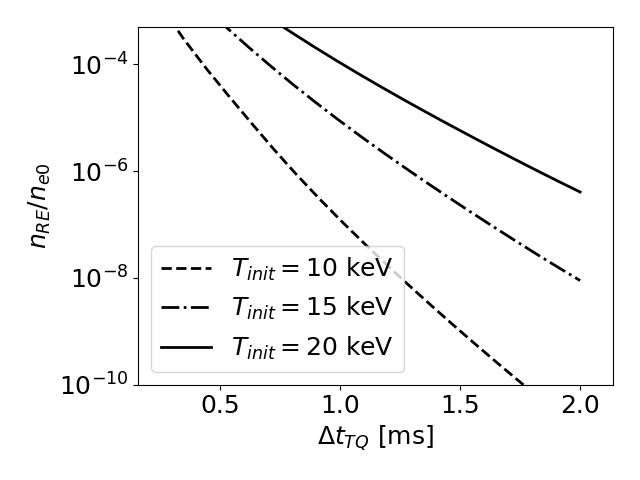}}
\subfigure[]{\includegraphics[scale=0.5]{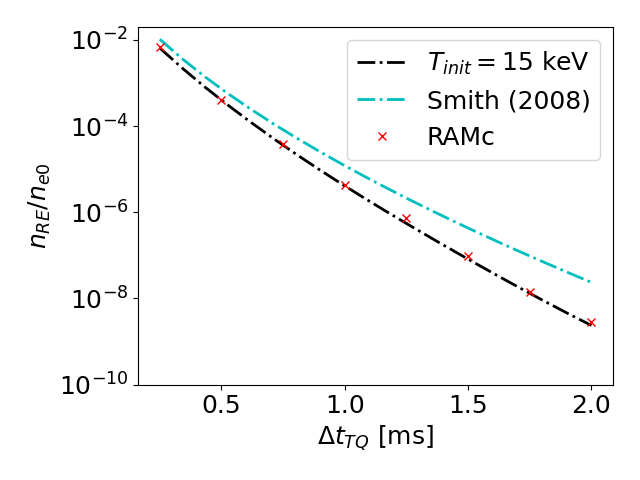}}
\par\end{centering}
\caption{The hot tail seed evaluated from a 6-D model as a function of the thermal quench time scale $\Delta t_{TQ}$ for different initial plasma temperatures $T_{init}$ and current densities $j_0$. Panel (a) is for $j_0=0.5\;\text{MA}/\text{m}^2$ and panel (b) is for $j_0=1\;\text{MA}/\text{m}^2$. Panel (c) shows a comparison with Eqs. (23) and (22) of Ref. \cite{smith2008hot}, for $T_{init}=15\;\text{keV}$ and $j_0=1\;\text{MA}/\text{m}^2$. The `x' markers indicate values found using the Monte Carlo code RAMc~\cite{mcdevitt2019avalanche}. The other parameters were taken to be $T_{final} = 30\;\text{eV}$ and $Z_{eff}=1$.}
\label{fig:GS2}
\end{figure}

In this section we will explore the broader parameter space of the idealized thermal quench model described above. In particular, we will train the model for a range of values for the parameters $\left( \Delta t_{TQ}, T_{init}, E_{init} \right)$ for fixed $T_{final}$. This latter parameter will be taken to be $T_{final}=30\;\text{eV}$, with the other three parameters trained within ranges of
$\Delta t_{TQ}$ between $0.25\;\text{ms}$ and $2 \;\text{ms}$, $T_{init}$ between $10\;\text{keV}$ and $20\;\text{keV}$, and the range of values of $E_{init}$ are chosen such that a plasma at a density of $n_e=10^{14}\;\text{cm}^{-3}$ will have plasma currents that range between $0.5\;\text{MA}/\text{m}^2$ and $1.5\;\text{MA}/\text{m}^2$. The loss history for the model is shown in Fig. \ref{fig:GS1}, with a minimum training loss of roughly $\approx 10^{-6}$ and a test loss of $\approx 2.4 \times 10^{-7}$. 
The hot tail seed evaluated as a function of $\Delta t_{TQ}$ for different initial temperatures and current densities is shown in Fig. \ref{fig:GS2}. Here, we have only included values of the hot tail seed less than $n_{RE}/n_{e0} = 5\times10^{-4}$ since for RE seeds larger than this the number of REs would be large enough to carry most, if not all, of the plasma current, thus violating the Ohm's law employed in Eq. (\ref{eq:TDRP16}). From Fig. \ref{fig:GS2} it is apparent that the magnitude of the hot tail seed is strongly sensitive to the initial temperature of the plasma $T_{init}$, with a somewhat weaker dependence on the current density $j_0$. The strong dependence on $T_{init}$ is largely due to the number of energetic electrons in the initial Maxwell-J\"{u}ttner distribution depending exponentially on $T_{init}$. In addition, $T_{init}$ impacts the electric field evolution through the $T_e$ dependence in Ohm's law [Eq. (\ref{eq:TDRP16})], though this dependence will have a far more modest impact on the number of hot tail REs generated. The current density $j_0$ (which enters via $E_0$) has a direct impact on the magnitude of the electric field. This influences the number of hot tail electrons via two distinct mechanisms. The first is that the magnitude of the electric field sets the critical energy to run away, which directly impacts the number of electrons that will be accelerated rather than slowed down. A second impact is that the electric field will also affect the form of the electron distribution function. This dependence will introduce additional anisotropy in the RPF, and thus have a more subtle impact on the hot tail mechanism.

\begin{figure}
\begin{centering}
\includegraphics[scale=0.5]{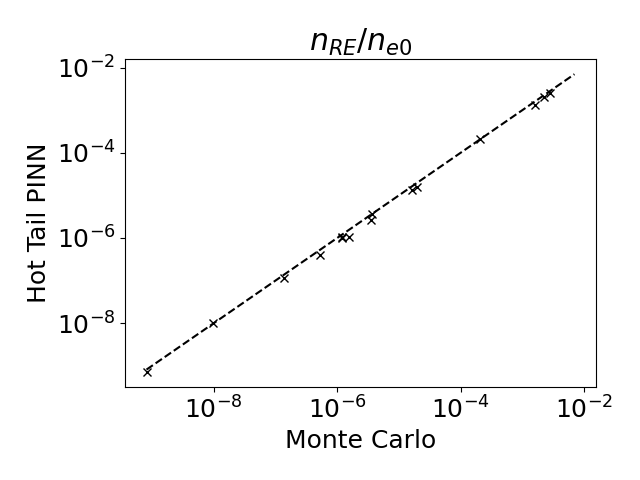}
\par\end{centering}
\caption{Comparison between the hot tail PINN and direct Monte Carlo simulations over a broad parameter space. $\left( \Delta t_{TQ}, T_{init}, j_0\right)$ were sampled randomly, with $T_{final} = 30\;\text{eV}$ for all cases.}
\label{fig:GS3}
\end{figure}

A comparison with Ref. \cite{smith2008hot} and first principle simulations of the hot tail mechanism using the RAMc code~\cite{mcdevitt2019avalanche} is shown in Fig. \ref{fig:GS2}(d). Here, the cyan curve indicates the approximate analytic theory defined by Eqs. (23) and (22) of Ref. \cite{smith2008hot}. From Fig. \ref{fig:GS2}(d) the hot tail PINN is in excellent agreement with the predictions from Monte Carlo simulations performed with RAMc, with the approximate analytic theory of Ref. \cite{smith2008hot} overestimating the magnitude of the hot tail seed. A comparison with Monte Carlo predictions across a broader range of parameters is shown in Fig. \ref{fig:GS3}. Here, we have considered a random sampling of the parameters $\left( \Delta t_{TQ}, T_{init}, j_0\right)$ across the range over which the hot tail PINN was trained. Excellent agreement is apparent between the direct Monte Carlo simulations and the predictions of the hot tail PINN,
suggesting the hot tail PINN was able to accurately learn the solution across the three-dimensional parameter space.

\begin{figure}
\begin{centering}
\subfigure[]{\includegraphics[scale=0.5]{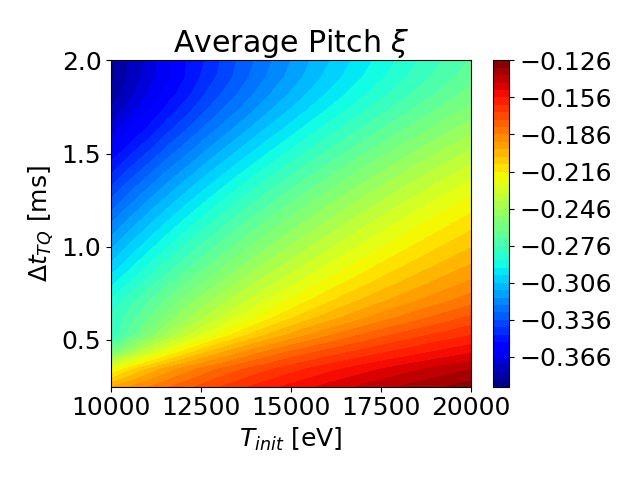}}
\subfigure[]{\includegraphics[scale=0.5]{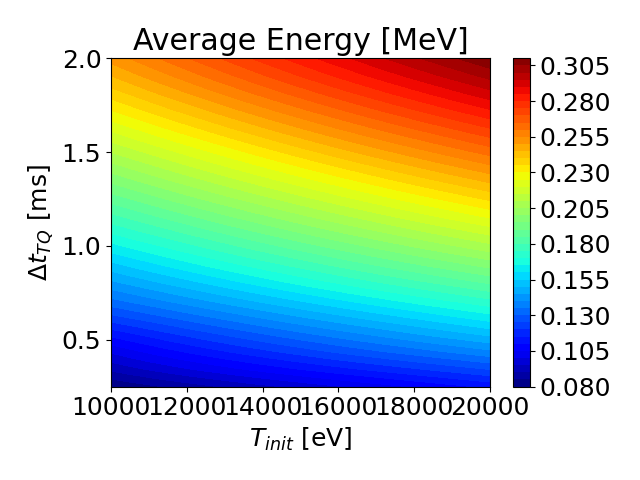}}
\par\end{centering}
\caption{Average initial pitch [panel (a)] and energy [panel (b)] of the hot tail electrons as a function of the initial temperature and thermal quench timescale. The plasma current was taken to be $j_0=1\;\text{MA/m}^2$.}
\label{fig:GS4}
\end{figure}

In addition to evaluating the number of seed electrons, the hot tail PINN can be used to infer additional properties of the electrons that make up the hot tail seed. Two properties that we will focus on here will be the average initial pitch and energy of the electrons that eventually run away. These can be computed by evaluating the moments
\begin{equation}
\left\langle \xi \right\rangle = \frac{1}{n_{RE}} \int d^3 p \xi f^{MJ}_e P^\prime \left( \tau=t_{final} \right) \chi \left( P^\prime \right)
, \label{eq:GS1}
\end{equation}
\begin{equation}
\left\langle E \right\rangle = \frac{1}{n_{RE}} \int d^3 p m_e c^2 \left( \gamma - 1 \right) f^{MJ}_e P^\prime \left( \tau=t_{final} \right) \chi \left( P^\prime \right)
. \label{eq:GS2}
\end{equation}
Considering the average pitch first, this quantity is shown as a function of initial temperature and thermal quench time in Fig. \ref{fig:GS4}(a) for an initial current density of $j_0= 1\;\text{MA}$. Here it is apparent that the electrons responsible for running away are on average moving in the negative pitch direction as expected, where this anisotropy becomes most acute for slow thermal quenches (long $\Delta t_{TQ}$) and low initial electron temperatures. Noting that several models of the hot tail neglect the anisotropy of the electron distribution, it is apparent that this approximation is never precisely satisfied, but breaks down quite severely for the case of a slow thermal quench. Turning to the average initial energy, here it is apparent that the typical initial energy of a hot tail RE ranges from slightly less than $100\;\text{keV}$, but can reach magnitudes up to roughly $200\;\text{keV}$, where the timescale of the thermal quench is the most important driver. At such high energies, a relativistic collision operator is required.

\section{\label{sec:DC}Discussion and Conclusions}

In the above, a PINN representation of the adjoint relativistic Fokker-Planck equation was developed in the presence of a rapidly evolving electron temperature and electric field. This hot tail PINN was used to evaluate the RE seed that emerges during an idealized axisymmetric thermal quench. The predictions from the hot tail PINN were shown to be in excellent agreement with direct solutions from the Monte Carlo code RAMc across a broad range of parameters, thus providing an effective surrogate model of the hot tail. Aside from the number of electrons predicted to run away, the PINN framework enabled additional properties such as the average initial pitch and energy of the electrons that would eventually run away to be extracted. This latter characteristic represents a strong point of the PINN framework, where rather than output a single quantity such as the runaway density, the entire RPF solution is encoded. This not only enables additional insight into properties of the solution, but also allows for greater interpretability of the predictions made by the hot tail PINN, and thus provides additional guidance as to whether a prediction of the surrogate model can be trusted.

While the present study provides a proof-of-principle demonstration of the potential of the hot tail PINN, 
several idealizations would need to be removed before a truly predictive model of the hot tail seed for a realistic tokamak disruption can be developed. Some of these idealizations would be conceptually simple to relax, such as treating a nonhydrogenic plasma or a more realistic temperature history, others would require significant extensions of the hot tail PINN developed above. Among the most challenging would be the inclusion of transport due to the partial or complete stochastization of the magnetic field arising from the presence of 3D MHD instabilities. While an ad-hoc spatial diffusivity due to MHD transport could be straightforwardly included in the above model, such an approach is unlikely to provide sufficient accuracy for a predictive hot tail model. Future work will instead seek to take advantage of the ease through which inverse problems may be treated within the PINN framework~\cite{cai2021physics, mathews2022deep}, with the aim of identifying a representative description of electron transport by a given spectrum of MHD modes. Once inferred, this transport model would be included in the hot tail PINN, where the properties of the MHD spectrum would correspond to additional parameters over which the hot tail PINN will be trained over.

\begin{acknowledgements}

This work was supported by the SciDAC project Tokamak Disruption Simulation (TDS) by Office of Fusion Energy Science and Office of Advanced Scientific Computing, and by the University of Florida Informatics Institute seed award. The authors acknowledge the University of Florida Research Computing for providing computational resources that have contributed to the research results reported in this publication.

\end{acknowledgements}

\end{document}